\documentclass[sigconf]{acmart}
\AtBeginDocument{%
  \providecommand\BibTeX{{%
    \normalfont B\kern-0.5em{\scshape i\kern-0.25em b}\kern-0.8em\TeX}}}

\copyrightyear{2024} 
\acmYear{2024} 
\setcopyright{rightsretained} 
\acmConference[WWW '24 Companion]{Companion Proceedings of the ACM Web Conference 2024}{May 13--17, 2024}{Singapore, Singapore}
\acmBooktitle{Companion Proceedings of the ACM Web Conference 2024 (WWW '24 Companion), May 13--17, 2024, Singapore, Singapore}
\acmDOI{10.1145/3589335.3651536}
\acmISBN{979-8-4007-0172-6/24/05}

\usepackage{multirow}
\usepackage[normalem]{ulem}
\useunder{\uline}{\ul}{}
\usepackage{bbm}
\usepackage{enumitem}
\usepackage{tablefootnote}
\usepackage{makecell}
\usepackage[ruled,linesnumbered,longend]{algorithm2e}
\usepackage{algpseudocode}  
\usepackage{amsmath}

\begin{document}

\title{MetroGNN: Metro Network Expansion with Reinforcement Learning}
  
\author{Hongyuan Su}
\affiliation{%
  \institution{Department of Electronic
Engineering, BNRist, 
Tsinghua University}
  \city{Beijing}
  \country{China}}

\author{Yu Zheng}
\affiliation{%
  \institution{Department of Electronic
Engineering, BNRist, 
Tsinghua University}
  \city{Beijing}
  \country{China}}

\author{Jingtao Ding}
\affiliation{%
  \institution{Department of Electronic
Engineering, BNRist, 
Tsinghua University}
  \city{Beijing}
  \country{China}}

\author{Depeng Jin}
\affiliation{%
  \institution{Department of Electronic
Engineering, BNRist, 
Tsinghua University}
  \city{Beijing}
  \country{China}}

\author{Yong Li}
\authornote{Corresponding author (liyong07@tsinghua.edu.cn).}
\affiliation{%
  \institution{Department of Electronic
Engineering, BNRist, 
Tsinghua University}
  \city{Beijing}
  \country{China}}

\renewcommand{\shortauthors}{Hongyuan Su, Yu Zheng, Jingtao Ding, Depeng Jin, \&Yong Li}
  
\begin{abstract}
Selecting urban regions for metro network expansion to meet maximal transportation demands is crucial for urban development, while computationally challenging to solve.
The expansion process relies not only on complicated features like urban demographics and origin-destination (OD) flow but is also constrained by the existing metro network and urban geography. 
In this paper, we introduce a reinforcement learning framework to address a Markov decision process within an urban heterogeneous multi-graph.
Our approach employs an attentive policy network that intelligently selects nodes based on information captured by a graph neural network.
Experiments on real-world urban data demonstrate that our proposed methodology substantially improve the satisfied transportation demands by over 30\% when compared with state-of-the-art methods.
Codes are published at \textcolor{blue}{\url{https://github.com/tsinghua-fib-lab/MetroGNN}}.

\end{abstract}




\maketitle
\section{Introduction}
Public transportation, especially the metro network, is a key component in meeting the travel needs of citizens, which not only helps to address the urban equity issue\cite{zheng2023road, liu2023knowsite}, but also influences the urban dynamics, including population distribution and economic development\cite{dong2019impact, zheng2023spatial}.
Therefore, a rational metro network expansion can have a profound impact on the development of the city.

Solving metro network expansion is nontrivial due to two primary challenges.
Firstly, consideration must be given to intricate features, including transportation flow matrices between regions and the relationship with existing metro lines\cite{driscoll2018effect}.
Secondly, the selection of regions for metro network expansion poses an NP-hard problem characterized by an enormous solution space comprising all candidate regions,  which makes it impossible to conduct an exhaustive search\cite{yolcu2019learning}.
For example, in a medium-sized city with 1000 regions, the solution space can exceed $10^{30}$, far beyond the capacity of exact solution methods. 
Additionally, the problem complexity is heightened by various constraints from urban geography\cite{bagloee2011transit}, including spacing and angles between stations and line segments.

Existing approaches for metro network expansion fall into three categories.
First, heuristics are proposed to select regions, such as greedy rules\citep{zarrinmehr2016path}, or genetic algorithms\citep{nayeem2018solving}.
Simulated annealing\citep{fan2006using} and ant colony methods\citep{yang2007parallel} are also adopted to better navigate the search space.
However, these heuristics struggle to handle the various constraints, leading to infeasible solutions.
Second, mathematical programming approaches eliminate the solution space by restricting the selected regions in narrow corridors\citep{wei2019strategic}.
Not surprisingly, the corridor approximation is oversimplified, which blocks out solutions of high quality and thus results in sub-optimal performance of the expanded metro network.
Reinforcement learning (RL) has recently been applied to this problem\citep{wei2020city}, but it neglects the consideration of diverse correlated features.

In this paper, we propose a systematic RL framework for solving complex Markov decision process (MDP) on a graph.
The metro network is expanded intelligently by selecting nodes on a heterogeneous multi-graph representing urban regions.
To address the challenge of complicated features, we design a novel graph neural network (GNN) to learn effective representations for the heterogeneous multi-graph.
Independent message propagation and neighbor aggregation are developed to capture both spatial contiguity and transportation flow between urban regions.
To efficiently explore the NP-hard problem solution space, we propose an attentive policy network with an action mask for region selection.
The action mask ensures the feasibility of the obtained solutions by addressing various metro network constraints.

To summarize, the contributions of this paper are as follows,
\begin{itemize}[leftmargin=*]
    \item We propose a graph-based RL framework for solving complex MDP, which is able to address the challenging geometrical CO problem of metro network expansion. 
    \item We design a novel GNN and an attentive policy network to learn representations for urban regions and select new metro stations.
    \item Extensive experiments on real-world urban data demonstrate that the proposed MetroGNN can substantially improve OD flow satisfaction by over 30\% against state-of-the-art approaches.
\end{itemize}

\section{Problem Statement}\label{sec::prob}
\begin{figure}[t]
    \centering
    \vspace{-5px}
    \includegraphics[height=0.3\linewidth,width=1.0\linewidth]{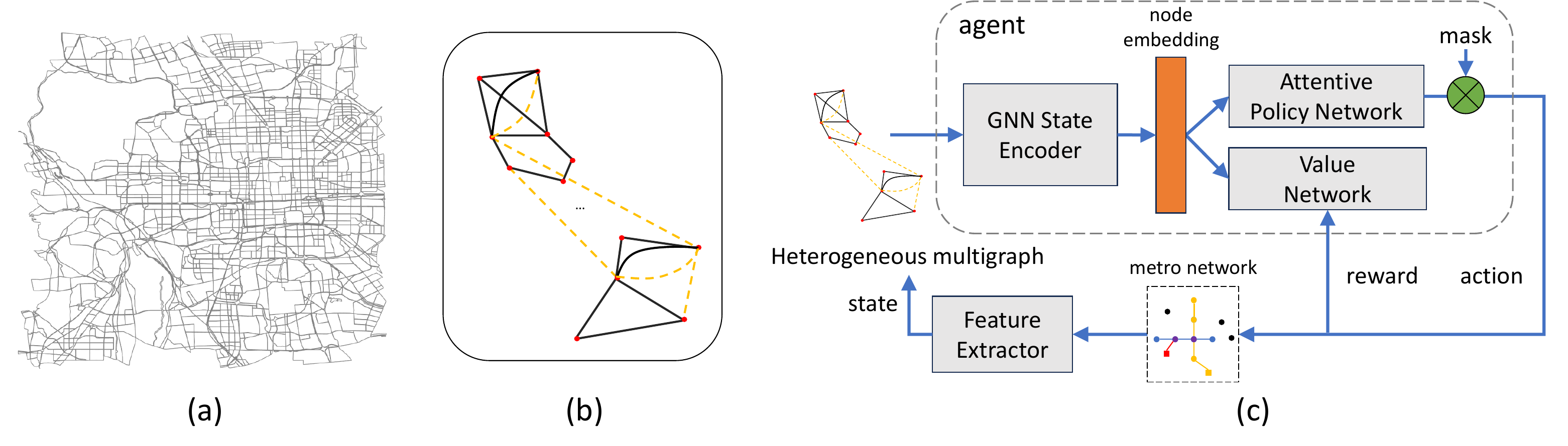}
    \vspace{-20px}
    \caption{
    (a) Regions determined by road network.
    (b) Heterogeneous multi-graph, where nodes represents regions.
    The black solid line and the orange dashed line correspond to spatial contiguity and OD associations between regions.
    (c) Schematic of our approach.
    }
    \vspace{-12px}
    \label{fig::graph}
\end{figure}

Given a set of nodes, $\mathcal{N}=\{n_1,n_2,...,n_k\}$, representing the centroids of urban regions (see Figure \ref{fig::graph}(a)), a metro network $\mathcal{M}=(\mathcal{V},\mathcal{E})$ can be described with a subset of nodes $\mathcal{V}\in\mathcal{N}$, and the edges $\mathcal{E}$ (metro lines segments) connecting nodes.
Metro network expansion involves the sequential selection of nodes for station construction to maximize its total satisfied OD flow, which can be quantified as follows:
\begin{align}\label{eq::od}
    C_{od}(\mathcal{M}) &= \sum_{(n_i,n_j) \in \mathcal{E}} \dfrac{\text{EucDis}(n_i,n_j)}{\text{PathDis}(n_i,n_j)} \cdot \mathcal{F}_{ij},
\end{align}

where $\text{EucDis}(n_i, n_j)$ and $\text{PathDis}(n_i, n_j)$ are the Euler distance and path distance between $n_i$ and $n_j$ via $\mathcal{M}$, respectively, and $\mathcal{F}_{ij}$ denotes the OD flow between $n_i$ and $n_j$.

\section{Method}
We propose a graph-based RL framework to solve the complex MDP with following components:
\begin{itemize}[leftmargin=*]
    \item \textbf{State.} The state $S_t$ is a two-tuple $(\mathcal{M}_t, b_t)$ containing the current metro network $\mathcal{M}_t=(\mathcal{V}_t,\mathcal{E}_t)$ and the remaining budget $b_t$.
    \item \textbf{Action.} The action $A_t$ corresponds to the selection of a single node in $\mathcal{N}$. 
    \item \textbf{Reward.} The intermediate reward $R_t$ for action $A_t$ is defined as ${C_{od}(\mathcal{M}_t)-C_{od}(\mathcal{M}_{t-1})}$.
    \item \textbf{State transition.} The selected node is regarded as an expansion of a metro line if there is no sharp bend; otherwise, it is considered as the start of a new line to expand the current metro network.
\end{itemize}

\subsection{Heterogeneous Multi-graph Model}\label{sec::graph_model}
\begin{figure}[t]
    \vspace{-5px}
    \centering
    \includegraphics[width=0.99\linewidth]{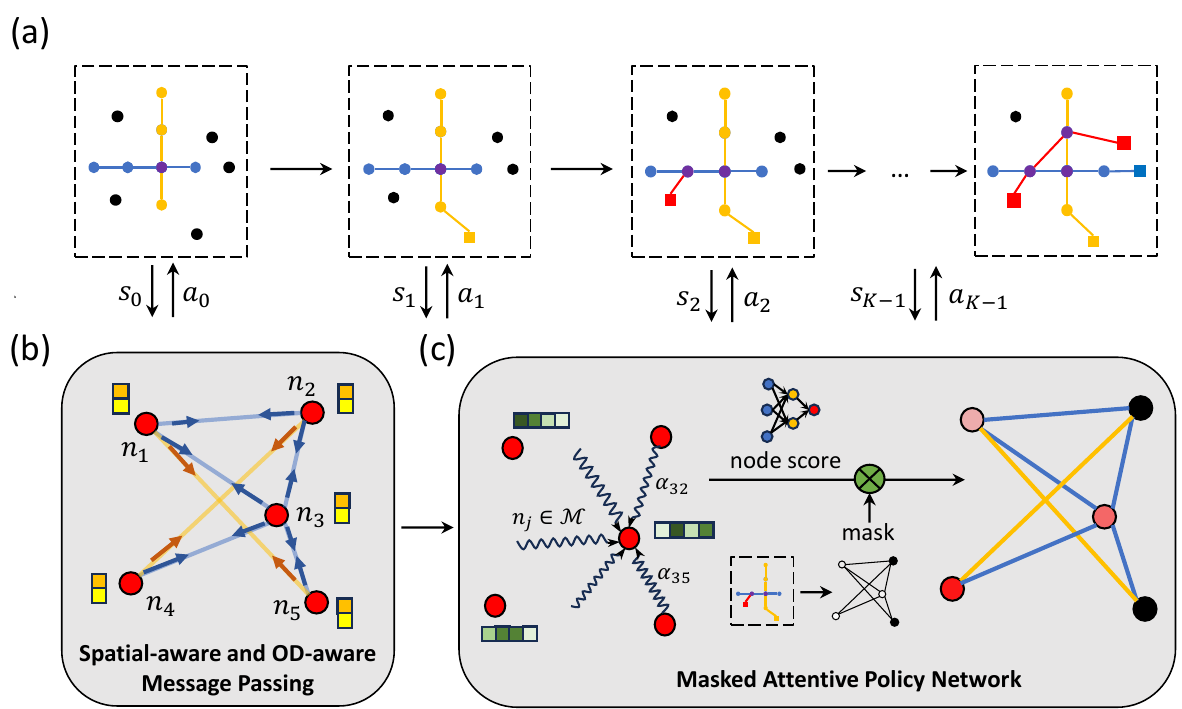}
    \caption{
    (a) The schematic of metro network expansion process. 
    At each step, the agent selects a node that either extends existing lines ($a_0$) or constructs new lines ($a_1$).
    We use distinct colors for different lines, and purple for interchange.
    (b) The proposed GNN model, where a spatial-aware and OD-aware message passing mechanism is developed.
    (c) The proposed masked attentive policy network for node selection.
    }
    \vspace{-12px}
    \label{fig::pipeline}
\end{figure}

As illustrated in Figure \ref{fig::pipeline}(a), We utilize a heterogeneous multi-graph to faithfully describe the urban regions.
In this graph model, the node set $\mathcal{N}=\{n_1,\cdots,n_k\}$ represents the regions divided by the road network.
We then introduce two types of edges to effectively capture the relationships between regions, as illustrated in Figure \ref{fig::graph}(b).
Specifically, the first type links contiguous nodes, capturing their proximity on small spatial scales.
The second type connects pairs of nodes with significant OD trips, capturing flow patterns between urban regions on a larger scale.
In conclusion, the heterogeneous edges are denoted as follows,
\begin{equation}
    e^s_{ij} = \mathbbm{1}\{0 < \text{EucDis}(n_i,n_j) 
    \leq{t_1}\},\quad e^o_{ij} = \mathbbm{1}\{\mathcal{F}_{ij} \geq {t_2}\},
    \label{eq::edge}
\end{equation}
where $t_1$ and $t_2$ are threshold values.
By expressing the problem with the graphical model, our framework can comprehensively express the spatial relationships and OD characteristics within city.

\begin{table*}[t]
\centering
\setlength{\tabcolsep}{3pt}
\renewcommand{\arraystretch}{1.1}
\begin{tabular}{c|ccc|ccc}
\toprule
\multirow{2}{*}{\textbf{Method}} & \multicolumn{3}{c|}{\textbf{Beijing}} & \multicolumn{3}{c}{\textbf{Changsha}}\\
                                 & \textbf{B=40} & \textbf{B=50} & \textbf{B=60}& \textbf{B=40} & \textbf{B=50} & \textbf{B=60}\\
\midrule
GS & 8.25$_{\pm 0.00}$ & 9.31$_{\pm 0.00}$ & 10.40$_{\pm 0.00}$ & 10.11$_{\pm 0.00}$ & 11.26$_{\pm 0.00}$ & 12.58$_{\pm 0.00}$\\
GA & 9.95$_{\pm 1.78}$ & 10.13$_{\pm 1.98}$ & 12.87$_{\pm 2.27}$ & 14.24$_{\pm 1.46}$ & 15.34$_{\pm 1.65}$ & 16.55$_{\pm 1.89}$\\
SA & 9.59$_{\pm 1.57}$ & 10.70$_{\pm 1.59}$ & 12.29$_{\pm 2.08}$ & 13.84$_{\pm 1.63}$ & 15.02$_{\pm 1.55}$ & 16.39$_{\pm 1.72}$\\
ACO & 11.01$_{\pm 1.14}$ & 12.42$_{\pm 1.30}$ & 13.66$_{\pm 1.45}$ & 16.61$_{\pm 1.59}$ & 17.67$_{\pm 1.52}$ & 17.17$_{\pm 1.97}$\\
MPC & 14.40$_{\pm 0.28}$ & 15.11$_{\pm 0.61}$ & 16.60$_{\pm 1.19}$ & 17.47$_{\pm 0.60}$ & 18.34$_{\pm 0.91}$ & \underline{20.43}$_{\pm 1.30}$\\
MPG & 14.40$_{\pm 0.28}$ & 15.16$_{\pm 0.93}$ & \underline{16.81}$_{\pm 1.13}$ & 17.47$_{\pm 0.91}$ & 18.07$_{\pm 1.26}$ & 20.17$_{\pm 1.63}$\\
DRL-CNN & \underline{14.46}$_{\pm 0.92}$ & \underline{15.78}$_{\pm 1.33}$ & 16.38$_{\pm 1.48}$ & \underline{18.30}$_{\pm 1.03}$ & \underline{18.98}$_{\pm 1.98}$ & 19.21$_{\pm 1.81}$\\
MetroGNN (ours) & \bf{15.88}$^{*}_{\pm 0.73}$ & \bf{18.93}$^{**}_{\pm 0.87}$ & \bf{21.45}$^{**}_{\pm 1.02}$ & \bf{20.79}$^*_{\pm 1.07}$ & \bf{22.72}$^{**}_{\pm 1.13}$ & \bf{24.65}$^{**}_{\pm 1.36}$\\
\hline
impr\% v.s. DRL-CNN & +9.8\% & +20.0\% & +31.0\% & +13.6\% & +19.7\% & +28.3\%\\
\toprule
\end{tabular}
\renewcommand{\arraystretch}{1.0}
\caption{
Evaluation of metro network expansion performance across varying budgets (\textbf{B}), higher is better. Statistical significance is determined using a t-test to compare MetroGNN with DRL-CNN, denoted as $^*\text{p-value}<0.1$ and $^{**}\text{p-value}<0.05$.
}
\label{tab::overall_1}
\vspace{-20px}
\end{table*}

\subsection{Encoding Complicated Features with GNN}\label{sec::gnn}
We design a novel GNN model as the encoder to learn unified representations of complicated features for regions through the heterogeneous edges, as shown in Figure \ref{fig::pipeline}(b).
Two groups of features are incorporated for each region.
The first group directly relates to the OD trips of the region, which includes the total OD access flows $FD_1$, the OD flows with neighboring regions $FD_2$ and the OD flows with the regions $\mathcal{V}$ where metro stations located $FD_3$.
The second group contains auxiliary features, including the population size $FA_1$, the type and number of Points of Interests (POIs) in each urban region $FA_2$, as well as topological features in the graphical model (including $FA_4$,$FA_5$ and $FA_6$).
The representation of nodes within the heterogeneous graph is computed as follows, 
\begin{align}
        &\mathbf{h}_i^{(0)} = \mathbf{W}_A\mathbf{A}_i, \\
        &\mathbf{h}^{(l)}_{s,i} = \sum_{j}e_{ij}^s \mathbf{W}_s^{(l)}\mathbf{h}^{(l)}_{j}, \quad \mathbf{h}^{(l)}_{o,i} = \sum_{j}e_{ij}^o \mathbf{W}_o^{(l)}\mathbf{h}^{(l)}_{j}, \\
        &\mathbf{h}_{i}^{{(l+1)}} = \tanh(\mathbf{W}_c^{(l)}(\mathbf{h}^{(l)}_{s,i} \,\|\, \mathbf{h}^{(l)}_{o,i}) + \mathbf{h}^{(l)}_{i}),
\end{align}
where $A_i$ is the input attribute for nodes, $\mathbf{W}_A$, $\mathbf{W}_s$, $\mathbf{W}_o$ and $\mathbf{W}_c$ are all linear layer and $\|$ denotes to concatenation.

With the proposed GNN model, we unify the complicated features for metro network expansion and obtain effective node representations with spatial contiguity and OD flow information.

\subsection{Planning with Masked Attentive Policy Network}
The solution space of metro network expansion expands exponentially with the number of expansion stations, making it exceedingly challenging to find optimal solutions, especially when considering various constraints such as straightness and spacing.
To search the massive solution space under various constraints, we propose an attentive policy network with an action mask for efficient exploration of feasible solutions, as illustrated in Figure \ref{fig::pipeline}(c).
The agent samples nodes according to the normalized scores, which is calculated by the proposed policy network based on embeddings computed by GNN as follows,
\begin{align}
    w_{ij} &= \frac{(\mathbf{W}_Q \mathbf{h}_{i}^{(L)})^T (\mathbf{W}_K \mathbf{h}_{j}^{(L)})}{\sqrt{d}}, \quad \alpha_{ij} = \frac{\exp(w_{ij})}{\sum_j \exp(w_{ij})}, \\
    \mathbf{a}_i &= \tanh\left(\sum_{v_j\in \mathcal{M}} \alpha_{ij} \cdot \mathbf{h}_j^{(L)} + \mathbf{h}_i^{(L)}\right), \quad \forall n_i \in \mathcal{N}, v_j \in \mathcal{V}, \\
    s_{i} &= \mathrm{MLP}_{p}(\mathbf{a}_{i}), \quad 
    p(n_i|\mathcal{M}) = \frac{\exp(s_{i})}{\sum_{i} \exp(s_{i})}e_{ij}^s .
\end{align}
where $d$ is the embedding dimension, $\mathbf{W}_Q, \mathbf{W}_K$ are learnable parameters, $w_{ij}$ is the relevance and $\alpha_{ij}$ is the attention score to the current metro network of each node $n_i$.
Nodes strongly correlated with the current metro network will be emphasized, while that away from the current metro network will be masked, enhancing high quality expansion of metro network.

\section{Experiments}
\subsection{Experiment Settings}

\noindent{\textbf{Data.}}
We conduct experiments using real-world data from two of China's largest cities, Beijing and Changsha.
Specifically, we adopt thousands of authentic urban region divisions delineated by the road structure.
Real OD flow data for the whole year of 2020 is utilized, which is obtained from Tencent Map, a prominent mapping and transportation service application in China.

\noindent{\textbf{Baselines and evaluation.}}
We compare our model with mathematical programming approaches~\citep{wei2019strategic} utilizing two different solvers, CBC (MPC) and GUROBI (MPG).
Heuristics baselines are also compared, including Greedy Strategy (GS)~\citep{laporte2015path}, Genetic Algorithm (GA)~\citep{owais2018complete}, Simulated Annealing Algorithm (SA)~\citep{fan2006using}, and Ant Colony Optimization (ACO)~\citep{yang2007parallel}.
We further include the state-of-the-art RL approach, DRL-CNN~\citep{wei2020city} for comparison.
For each method, we vary the seeds and conduct each experimental configuration 10 times.
To evaluate the effect of metro network expansion, we calculate the OD flows satisfied by the expanded metro network according to (\ref{eq::od}).

\subsection{Performance Comparison}\label{sec::perf_comp}
We assess each method under different scenarios by varying the total budget for expansion and evaluating the corresponding performance.
Results of our model and baselines are illustrated in Table \ref{tab::overall_1} and we have the following observations,
\begin{itemize}[leftmargin=*]
    \item \textbf{DRL-based methods have significant advantages over other approaches.}
    DRL-CNN outperforms other baselines in most cases, achieving higher satisfied OD with an average improvement of 4.4\%, demonstrating the superior ability of RL to search a large solution space.
    Nevertheless, DRL-CNN suffers from severe performance deterioration in complicated scenarios (B=60), with the satisfied OD 5.1\% worse than MPC and MPG on average.
    \item \textbf{Our proposed model achieves the best performance in different scenarios.}
    Our approach substantially surpasses existing baselines under all budgets, substantially improving the satisfied OD flow by over 15.9\% against the best baseline in average of three different expansion budgets.
    Notably, in contrast to DRL-CNN that fails to outperform baselines in complicated scenarios, our approach exhibits more significant advantages in complicated scenarios with a higher budget, with improvements on satisfied OD even over 30\%.
\end{itemize}

\begin{figure}[t]
    \centering
    \includegraphics[width=0.99\linewidth]{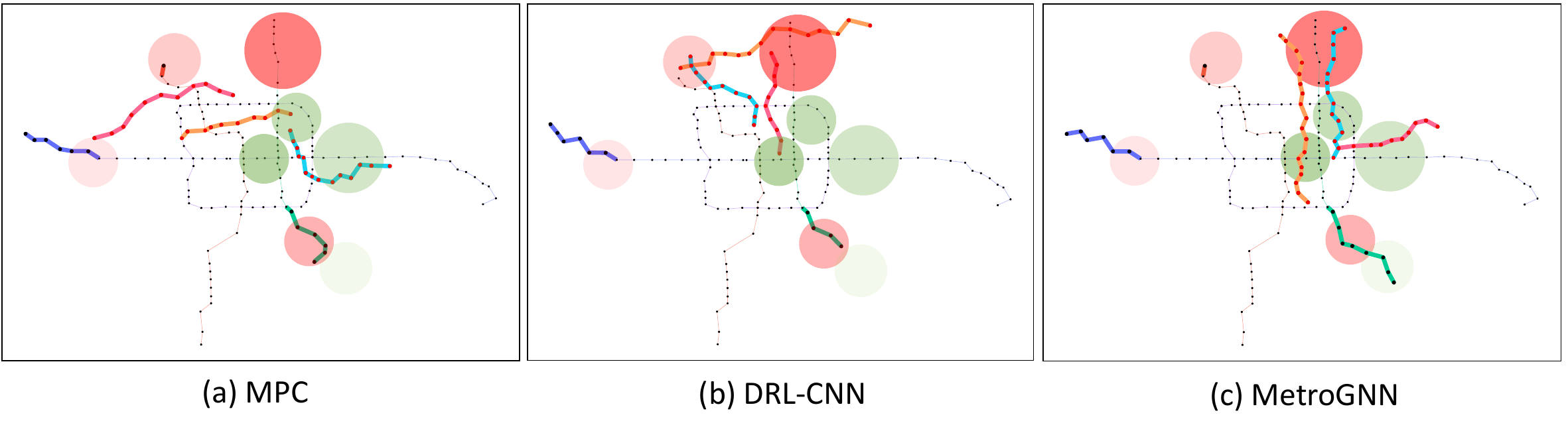}
    \vspace{-12px}
    \caption{
    Visualization of metro network expansion for Beijing.
    We use colors to distinguish between different metro lines, and use boldface nodes to indicate expansion solution.
    Black nodes indicate stations on the initial lines and their extensions, and red nodes represent stations on new lines.
    Regions colored with red and green indicate areas where population and POIs are clustered, respectively.
    The darker the color, the higher the density.
    }
    \label{fig::mainresult}
    \vspace{-15px}
\end{figure}

To provide a deeper understanding of the reliability and practical applicability of our planning solution within real-world contexts, we present the results generated by the MetroGNN and other baselines based on Beijing, as illustrated in Figure \ref{fig::mainresult}.
The planning solution generated by our approach covers almost all the areas with high population and POI densities, which naturally correspond to numerous travel demands.
Meanwhile, the new lines generated by our approach are interconnected, and each new line introduces at least two additional interchange stations to the metro network, improving the efficiency of the transportation network.

\subsection{Ablation Study}\label{sec::ablation}
We conduct ablation experiments to showcase the efficacy of the graph model and the incorporated complicated features.

\begin{figure}[t]
    \centering
    \vspace{-0px}
    \includegraphics[width=0.99\linewidth]{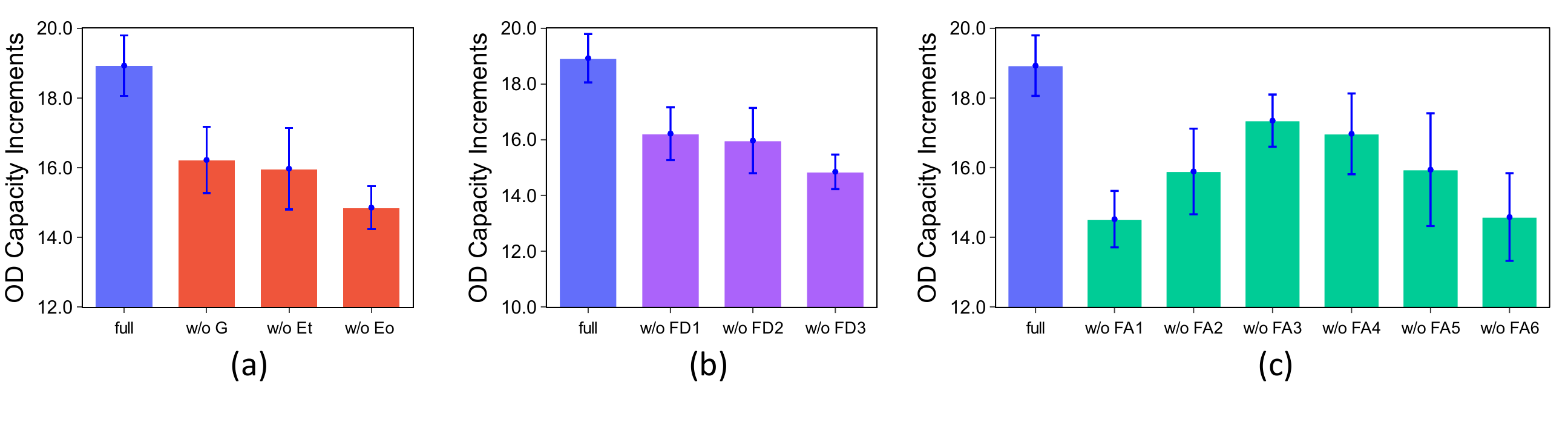}
    \vspace{-15px}
    \caption{
    Performance of MetroGNN and its variants that remove different elements, including whole graph model (G), spatial edges (Et), transportation flow edges (Eo), OD direct (FD) and auxiliary (FA) features.
    Best viewed in color.
    }
    \vspace{-15px}
    \label{fig::ablation}
\end{figure}

\noindent\textbf{Graph Modeling.}
The urban regions exhibit intricate spatial correlations characterized by both short-range proximity and long-range OD flow patterns.
By harnessing the graph modeling approach and GNN, our approach effectively captures these complexities among urban regions.
As shown in Figure \ref{fig::ablation}(a), when the graph model is omitted, the satisfied OD flow of the expanded metro network drops significantly from 21.80 to only 14.02.

\noindent\textbf{Spatial-aware and OD-aware Message Passing.}
In the proposed GNN model, we design two independent message propagation mechanisms, spatial-aware and OD-aware message passing.
As illustrated in Figure \ref{fig::ablation}(a), removing either spatial or transportation edges leads to a significant deterioration in performance, with a decrease of 28.8\% and 30.8\%, respectively.

\noindent\textbf{OD Direct and Auxiliary Features.}
As shown in Figure \ref{fig::ablation}(b), when the three OD direct features are excluded, our method observes varying degrees of performance degradation.
In particular, removing FD3 results in the largest performance drop (-20.4\%), which is reasonable since it reflects the direct benefit of adding a region to the metro network.
We also evaluate the contribution of auxiliary features, as demonstrated in Figure \ref{fig::ablation}(c).
In particular, removing population (FA1) brings about the largest deterioration, as population information is quite important when considering metro network expansion.

\section{Conclusion}
In this paper, we investigate the problem of metro network expansion, and propose MetroGNN, a systematic graph-based RL framework that can solve complex node selection MDPs on the graph. 
The proposed model unifies complicated features with GNN and explores the solution space efficiently with an attentive policy network and a carefully designed action mask.
Through extensive experiments, our approach demonstrates a significant improvement on travel demand satisfaction, increasing the satisfied OD flow by over 15.9\% compared to state-of-the-art baselines.
Looking ahead, we plan to investigate the performance of the proposed systematic RL framework in other graph-based decision tasks, such as influence maximization on social media platforms.

\begin{acks}
This work is supported in part by The National Natural Science Foundation of China under U20B2060, U22B2057, 62272260. This work is also supported in part by Beijing National Research Center for Information Science and Technology (BNRist).
\end{acks}

\bibliographystyle{ACM-Reference-Format}
\bibliography{references}

\end{document}